\documentclass[aps,prb,reprint,groupedaddress]{revtex4-1}

\usepackage{amssymb}
\usepackage{amsmath}
\usepackage{graphicx}
\usepackage{epsfig}

\begin{document}

\title{Soliton induced critical current oscillations in two-band superconducting bridges}

\author{P.M. Marychev}
\email[Corresponding author: ]{marychevpm@ipmras.ru}
\author{D.Yu. Vodolazov}
\affiliation{Institute for Physics of Microstructures, Russian
Academy of Sciences, Nizhny Novgorod, 603950 Russia}

\date{\today}

\begin{abstract}

Using time-dependent Ginzburg-Landau theory we find oscillations
of critical current density $j_c$ as a function of the length $L$
of the bridge formed from two-band superconductor. We explain this
effect by appearance of the phase solitons in the bridge at
$j<j_c$, those number changes with change of $L$. In case of
sufficiently strong interband coupling oscillations of $j_c$
disappear.
\end{abstract}

\maketitle

\section{Introduction}

The discovery of two-band superconductivity in
MgB$_2$~\cite{Nature-2002} has caused an explosion of research on
physical properties of such materials. Later the evidence of two
and more superconducting gaps has been observed in many materials,
e.g. in OsB$_2$,~\cite{PRB-2010} NbSe$_2$,~\cite{PRL-2003}
LiFeAs,~\cite{PRB-2011} FeSe$_{0.94}$~\cite{PRL-2010} and other
iron-based superconductors. Theoretically superconductivity in
each band can be described by complex order parameter
$\Psi_k=\Delta_k\exp(i\theta_k)$, where $\Delta_k$ and $\theta_k$
are superconducting amplitude and phase in k-th band respectively.
In the case of negligible interband impurity scattering, interband
interaction can be described by Josephson coupling
$\gamma_{ij}\Delta_i\Delta_j \cos(\theta_i-\theta_j)$ in
Ginzburg-Landau free energy functional. Interband phase difference
$\theta_i-\theta_j$ is determined by the sign of Josephson
coupling constant $\gamma_{ij}$ and can be either 0 or $\pi$ (so
called phase-locked state). However, in addition to the
phase-locked states, the existence of phase solitons is possible,
first discussed by Tanaka.~\cite{PRL-2002} In this states the
distribution $\theta_i-\theta_j$ is spatially inhomogeneous and
rotates by $2\pi$. Such objects have been observed in artificial
multiband structure consisting of mesoscopic aluminium
rings~\cite{PRL-2006} and also there is the evidence of their
existence, obtained in the experiment with a cuprate
film.~\cite{SSC-2015}

The phase solitons generally are not present in the ground state
and various methods to create them have been proposed. Gurevich
and Vinokur~\cite{PRL-GV-2003} used to excite phase solitons in
two-band superconductors by an electric field, where the boundary
between superconducting wire and normal lead was the solitons
source. It is also possible to create phase solitons at
equilibrium when the supercurrent in the weak band (band with
smaller superconducting amplitude) exceeds the critical value. It
results in phase slippages in weak band, which in turn generate
the chain of phase solitons. \cite{PRL-GV-2006}

In our work in the framework of one-dimensional time-dependent
Ginzburg-Landau theory we investigate influence of interband
breakdown (transition from phase-locked to soliton state) on
critical current density $j_c$, at which stationary
superconducting state ceases to exist in a two-band
superconducting bridge. \cite{self} Unlike the authors of
work,~\cite{PRL-GV-2006} we consider influence of strength of
Josephson coupling $\gamma$ and bridge length on $j_c$. This
breakdown is specific to systems with weak or absent interband
Josephson coupling and arises when supervelocity exceeds critical
supervelocity $q_{c2}$ in the weak band. In the case zero and
small $\gamma$ for length $L>>\xi_1$ ($\xi_1$ -- is the coherence
length in the strong band), there are oscillations of critical
current as a function of the bridge length, related to the change
of number of phase solitons in the system. Also there is a range
of parameters, in which the dependence $j_c(L)$ has minimum,
corresponding to transition from the phase soliton to the
phase-locked state. If interband coupling is sufficiently strong,
phase soliton generation is suppressed and above-mentioned
features are absent.

\section{Method}

To study the current states and find the critical current of
superconducting bridge, we numerically solve one-dimensional
time-dependent Ginzburg-Landau equation (TDGL)~\cite{Kopnin}
generalized for a two-band superconductor \cite{PRL-GV-2003} and
coupled with the equation for total current density $j$ in the
bridge. In the dimensionless form these equations are written as

\begin{widetext}
\begin{eqnarray}
 \label{TDGL-eq-1-dimless}
  u_1\left(\frac{\partial}{\partial
  t}+i\varphi\right)\Psi_1-\Psi_1+|\Psi_1|^2\Psi_1-\frac{\partial^2\Psi_1}{\partial x^2}-\gamma\Psi_2=0,
  \\
   \label{TDGL-eq-2-dimless}
    u_2\left(\frac{\partial}{\partial
  t}+i\varphi\right)\Psi_2-\alpha\Psi_2+\beta|\Psi_2|^2\Psi_2-g\frac{\partial^2\Psi_2}{\partial x^2}-\gamma\Psi_1=0,
  \\
   \label{current-dimless}
   j=-(1+\sigma)\frac{\partial\varphi}{\partial x}+
   Im\left(\Psi^*_1\frac{\partial\Psi_1}{\partial x}\right)+g
    Im\left(\Psi^*_2\frac{\partial\Psi_2}{\partial x}\right),
\end{eqnarray}
\end{widetext}
where $\alpha=|\alpha_2|/|\alpha_1|$, $\beta=\beta_2/\beta_1$,
$g=g_2/g_1$ and $\sigma=\sigma_{n2}/\sigma_{n1}$. Here $\varphi$
is the electric potential, $\sigma_{nk}$ is normal conductivity in
k-th band. $\alpha_{k}$, $\beta_{k}$ and $g_{k}$ are the GL
expansion coefficients, $\gamma$ is the interband Josephson
coupling constant. $\alpha_{k}$, $\beta_{k}$, $g_{k}$ and $\gamma$
was derived within the BCS theory through microscopic parameters
in Refs. ~\cite{PRB-2004} and~\cite{PRB-2003} for clean and dirty
superconductors respectively. Coordinate $x$ is measured in units
of the coherence length in the first (strong) band
$\xi_1=\sqrt{g_1\hbar^2/|\alpha_1|}$, order parameters are
measured in units of $|\Psi_{10}|=|\alpha_1|/\beta_1$, time is
measured in units of the current relaxation time in the first band
$t_{1}=\beta_1\sigma_{n1}/8e^2g_1|\alpha_1|$, Josephson coupling
constant $\gamma$ is in units of $|\alpha_1|$, and the current
density is measured in units of $j_0=4e\hbar g_1
|\Psi_{10}|^2/\xi_1$. In the phase-locked state interband phase
difference $\theta=\theta_1-\theta_2$ is determined by the sign of
interband coupling constant $\gamma$: for $\gamma>0$ phase
difference $\theta=0$ and $\theta=\pi$ for $\gamma<0$. Further we
consider case $\gamma>0$.

Parameters $u_k$ in
Eqs.~\eqref{TDGL-eq-1-dimless},\eqref{TDGL-eq-2-dimless} describe
the ratio of relaxation times of the order parameter in the
corresponding band to the current relaxation time in the first
band. In the case of gapless superconductors parameter
$u_{1,2}=5.79$ \onlinecite{UFN-1984}. For finite gap
superconductors one still can use Eqs.
\eqref{TDGL-eq-1-dimless},\eqref{TDGL-eq-2-dimless} but with
larger value of $u_k$ to take into account relatively large
relaxation time of $|\Psi_k|$ due to inelastic electron-phonon
scattering. \cite{Vodolazov-2011} As we show later large $u_k$
could lead to smaller value of the critical current but our main
result - oscillations of the critical current with changing length
of the bridge does not depend on specific value of $u_k$.
Therefore in calculations we use as small $u_k<1$ (providing
largest value of $j_c$) as large $u_k >> 1$. For simplicity and in
order to not multiply the value of independent parameters we
choose $u_1=u_2$.

Equations \eqref{TDGL-eq-1-dimless}---\eqref{current-dimless} are
solved for bridges with superconducting and normal leads. We
assume that the cross section of superconducting leads is much
larger than the cross section of the bridge. Then the equations
\eqref{TDGL-eq-1-dimless}---\eqref{TDGL-eq-2-dimless} are solved
with the boundary condition

\begin{eqnarray}
 \label{psi-scs-bound}
\Psi_{k}(0,t+\Delta
t)=\Psi_{k}(0,t)=|\Psi^0_{k}|,
 \\
 \nonumber
\Psi_{k}(L,t+\Delta t)=\Psi_{k}(L,t)e^{-i\varphi_L\Delta t},
\end{eqnarray}
where $\Psi^0_{k}$ are bulk order parameters at zero current, $L$
is the bridge length, $\varphi_L$ is the electric potential at the
point $x=L$ and $\Delta t$ is the time step. To find electric
potential $\varphi$ the equation \eqref{current-dimless} is solved
for each time step with given total current and the boundary
condition $\varphi (0)=0$. We use following criterion for reaching
the stationary state: $\varphi (L)-\varphi (0)=0$.

In the case of normal leads we use boundary conditions

\begin{equation}
 \label{psi-sns-bound}
\Psi_{k}(0,t)=\Psi_{k}(L,t)=0.
\end{equation}
with following criterion for the stationary state:
$\partial|\Psi_{k}|/\partial t=0$.

\section{Current states in two-band superconductor without interband Josephson coupling}
\label{sec:decoupled}

First we consider the case of zero interband Josephson coupling
$\gamma=0$. Such two-band system corresponds to either liquid
metallic hydrogen~\cite{PRL-1968} or superconductor with
negligible interband coupling $\gamma<<\alpha$. In the ground
state interband phase difference $\theta=0$ which is ensured by
boundary conditions and the absence of electric field in the
bridge. For long bridge Eqs.
\eqref{TDGL-eq-1-dimless}---\eqref{current-dimless} can be written
in the stationary case as

\begin{eqnarray}
 \label{GL-eq-1-locked}
 |\Psi_{1}|((q^{pl})^2-1)+|\Psi_{1}|^3=0,
 \\
 \label{GL-eq-2-locked}
  |\Psi_{2}|((q^{pl})^2-\alpha)+\beta|\Psi_{2}|^3=0,
  \\
  \label{current-locked}
  j=q^{pl}(|\Psi_{1}|^2+g|\Psi_{2}|^2),
\end{eqnarray}
where $q^{pl}=\partial\theta_1/\partial
x=\partial\theta_2/\partial x$ is supervelocity in the
phase-locked state. For simplicity we assume $\beta=g=1$. Using
these equations, one can obtain expressions for the supervelocity,
critical supervelocity and supercurrent in the phase-locked state,
when supervelocities in both bands are the same

\begin{widetext}
\begin{eqnarray}
 \label{supervelocity-locked}
  q^{pl}=2\sqrt{\frac{1+\alpha}{6}}\cos\left[\frac{1}{3}\arccos\left(-\frac{j}{4}\left(\frac{6}{1+\alpha}\right)^{3/2}\right)-\frac{2\pi}{3}\right],
  \\
  \label{qc-locked}
   q^{pl}_c=\sqrt{\frac{1+\alpha}{6}},
   \\
  \label{depairing-locked}
  j^{pl}_c=\frac{2}{3}\sqrt{\frac{(1+\alpha)^3}{6}}.
\end{eqnarray}
\end{widetext}

In the case of different order parameters, i.e. $\alpha<1$, the
critical supervelocity $q^{pl}_c$ is larger than the critical
supervelocity in the second band $q_{c2}=\sqrt{\alpha/3}$. Then,
at $q^{pl}>q_{c2}$ superconductivity is destroyed in the weak
band, and maximum supercurrent in phase-locked state is limited by
$j(q_{c2})$. Arising phase slip process and electric field
redistributes supervelocities, which in turn causes current
redistribution between bands. As a result, the system goes into
stationary, uniform state with linear interband phase difference
$\theta=2\pi mx/L$ (soliton-like state with size of soliton core
equal to length of the bridge), where $m$ is an integer number
which is equal to difference between number of phase slips in the
strong and weak bands (it is equal to number of solitons in the
bridge). If we choose $\theta(x=0) =0$ then at $x=L$ interband
phase difference $\theta =2\pi m$, i.e. the phase-locked state in
the leads is preserved. In this soliton state the order parameters
are defined by usual formulas $|\Psi_{1}|=\sqrt{1-q^2_1}$,
$|\Psi_{2}|=\sqrt{\alpha-q^2_2}$, and supervelocities
$q^{ps}_{1,2}$ with supercurrent are determined by following
expressions

\begin{widetext}
\begin{eqnarray}
 \label{supervelocity-ps}
  q^{ps}_1=2\sqrt{\frac{1+\alpha}{6}-\frac{\pi^2m^2}{L^2}}\cos\left[\frac{1}{3}\arccos\left(-\frac{\left(\frac{\pi
  m}{L}(\alpha-1)+j\right)\left(\frac{1+\alpha}{6}-\frac{\pi^2m^2}{L^2}\right)^{-3/2}}{4}\right)-\frac{2\pi}{3}\right]+\frac{\pi
  m}{L},
  \\
  \nonumber
   q^{ps}_2=2\sqrt{\frac{1+\alpha}{6}-\frac{\pi^2m^2}{L^2}}\cos\left[\frac{1}{3}\arccos\left(-\frac{\left(\frac{\pi
  m}{L}(\alpha-1)+j\right)\left(\frac{1+\alpha}{6}-\frac{\pi^2m^2}{L^2}\right)^{-3/2}}{4}\right)-\frac{2\pi}{3}\right]-\frac{\pi
  m}{L},
   \\
   \label{criticalsupervelocity-ps}
   q^{ps}_{c1}=\sqrt{\frac{1+\alpha}{6}-\frac{\pi^2m^2}{L^2}}+\frac{\pi
  m}{L}, q^{ps}_{c2}=\sqrt{\frac{1+\alpha}{6}-\frac{\pi^2m^2}{L^2}}-\frac{\pi
  m}{L},
   \\
   \label{current-ps}
   j^{ps}_c=q^{ps}_{c1}(1-(q^{ps}_{c1})^2)+q^{ps}_{c2}(\alpha-(q^{ps}_{c2})^2).
\end{eqnarray}
\end{widetext}
The value of $m$ is determined by conditions $q^{ps}_1<q_{c1}$,
$q^{ps}_2<q_{c2}$ in the stationary state and length of the bridge
because each phase slip changes the phase difference in each band
along the bridge by $2 \pi$ leading to corresponding change of
supervelocities $q_{1,2}$ by $\sim \pm 2\pi/L$. Since the
expressions \eqref{criticalsupervelocity-ps} depend on the bridge
length too, we can expect oscillations of critical current with
increasing the length of the bridge, which corresponds to change
in the number of phase slips and phase solitons.

\begin{figure}
\includegraphics[width=1.0\linewidth]{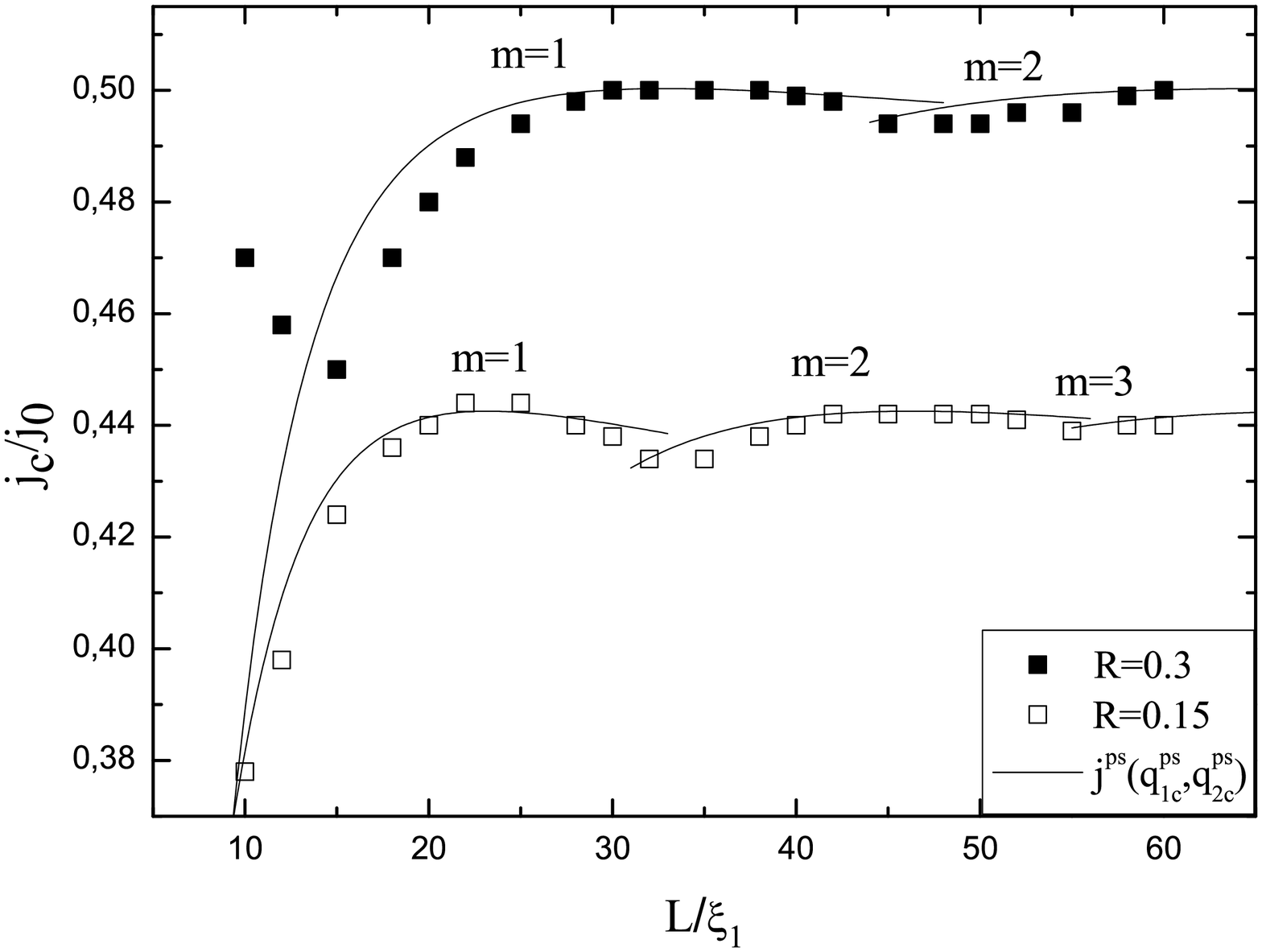}
 \caption{\label{Fig:jc-uncoup-long}
The critical current $j_c$ as a function of the bridge length $L$
in the absence of interband Josephson coupling at different values
of parameter $R=0.15$, $R=0.3$ and $u_1=u_2=0.2$. The solid lines
correspond to Eq. (14) with different $m$.}
\end{figure}

Fig.~\ref{Fig:jc-uncoup-long} shows the comparison of numerically
calculated critical current $j_c$ with analytical expression
$j^{ps}(q^{ps}_{c1},q^{ps}_{c2})$ (see Eq. (14)). We choose the
ratio of critical currents of different bands (superconductors)
$R=j_{c2}/j_{c1}$ as a parameter reflecting experimentally
observed difference between the band characteristics (or, in the
case of artificial structure, characteristics of superconductors).
Minimums of the dependence $j_c(L)$ correspond to crossover
between regions with different number of solitons.

\begin{figure}
\includegraphics[width=1.0\linewidth]{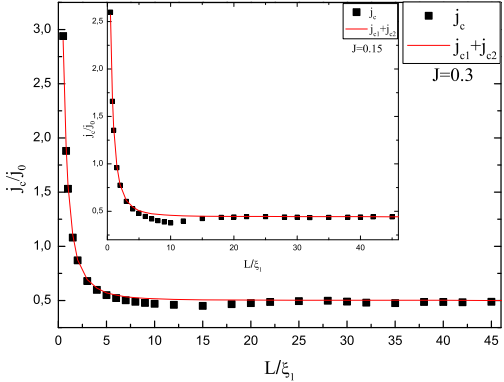}
 \caption{\label{Fig:jc-uncoup}
The critical current density $j_c(L)$ in the absence of interband
Josephson coupling at different values of parameter $R$ and
$u_1=u_2=0.2$. Solid curve corresponds to the sum of intraband
critical currents $j_{c1}+j_{c2}$.}
\end{figure}

At $m=1$ critical current $j_c$ decreases with decreasing length
$L$, according to formulas \eqref{criticalsupervelocity-ps} è
\eqref{current-ps}. For relatively small bridge length $L\lesssim
10\xi_1$ supervelocity redistribution caused by phase slips is too
large to keep the system in stationary state and up to critical
current the bridge stays in the phase-locked state. For such a
small lengths $j_c \sim 1/L$ as in single band supercondusting
bridge.~\cite{JETPL-1969} Therefore, the dependence $j_c(L)$ has a
minimum at $L \sim 10-20\xi_1$, which corresponds to the crossover
between the phase-locked state (small lengths) and one soliton
state (large lengths). Because position of the minima is
associated with value $j^{pl}_c$, it is very sensitive to values
of $\alpha$ and $j_{c2}$, see Eq. \eqref{depairing-locked}. This
feature is confirmed by numerical calculations (see
Fig.~\ref{Fig:jc-uncoup}).

We also did calculations for bridge with normal leads. In this
case oscillations $j_c(L)$ and the minimum at $L\sim 10-20\xi_1$
are absent. We explain it by absence of rigid boundary conditions,
i.e. there is no fixed interband phase difference $\theta (L)
=2\pi m$.

\section{Case of finite interband coupling}

Let us consider now how finite interband Josephson coupling
influences the critical current of superconducting bridge. As it
is shown in Ref.\onlinecite{PRL-GV-2006}, presence of weak
Josephson coupling leads to appearance of the phase solitons at
$q^{pl}>q_{c2}$. Additionally, in the work~\cite{NJP-2012} it is
shown that phase solitons are stable only for small interband
coupling constant $\gamma$. Therefore one could expect that the
phase solitons can affect the critical current only for small
$\gamma<<\alpha_1,\alpha_2$.

\begin{figure}
\includegraphics[width=1.0\linewidth]{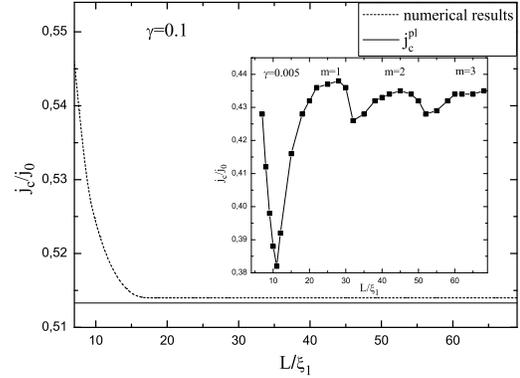}
 \caption{\label{Fig:jc-coupled}
The critical current density $j_c(L)$ for the interband Josephson
coupling $\gamma=0.005$ (see inset) and $\gamma=0.1$ with
$R=0.15$. Calculations are done for the value of parameter
$u=u_1=u_2=0.2$. Solid curve is Eq. (14).}
\end{figure}

\begin{figure}
\includegraphics[width=1.0\linewidth]{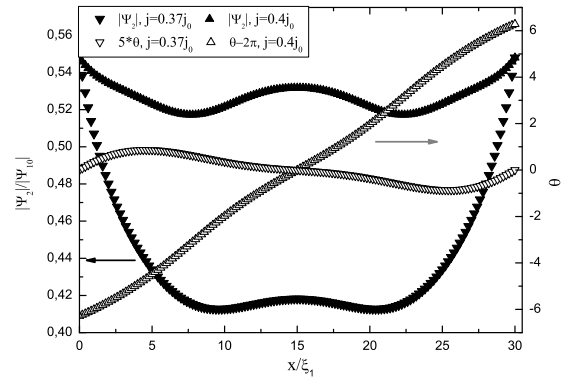}
 \caption{\label{Fig:psi-theta}
The amplitude of second gap $|\Psi_2|$ (black triangles) and
interband phase difference $\theta$ (white triangles) for the
interband Josephson coupling $\gamma=0.01$, the parameter
$u=u_1=u_2=0.5$, $R=0.15$ and the current densities $j=0.37j_0$
(down triangles) and $j=0.4j_0$ (up triangles). For illustrative
purposes the interband phase difference for the current density
$j=0.37j_0$ is multiplied by five and the interband phase
difference for the current density $j=0.4j_0$ is shifted down by
$2\pi$. }
\end{figure}

To find the critical current $j_c$ we numerically solve Eqs.
\eqref{TDGL-eq-1-dimless}---\eqref{current-dimless} for different
bridge lengths, different parameters $\gamma$ and we mainly use
$R=0.15$. The results for $\gamma=0.005$ and $\gamma=0.1$ are
shown in Fig.~\ref{Fig:jc-coupled}. When $\gamma$ is small the
phase-locked state is destroyed in a way similar to the case
$\gamma=0$. The only difference is that the presence of finite
$\gamma$ gives rise to nonlinear dependence $\theta (x)$ along the
bridge which also lead to small variations of $|\Psi_k|(x)$
(see~Fig. \ref{Fig:psi-theta}). As one can see from inset to
Fig.~\ref{Fig:jc-coupled}, oscillations of critical current
density $j_c(L)$ corresponds to changing the number of phase
solitons $m$ as length of the bridge changes. Crossover between
regions with different $m$ are responsible for the minimums of
$j_c(L)$.

\begin{figure}
\includegraphics[width=1.0\linewidth]{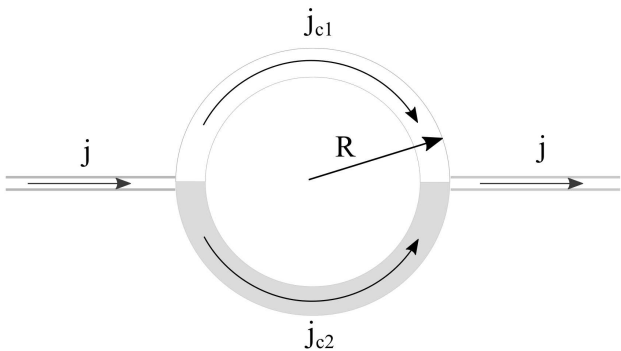}
 \caption{\label{Fig:ring}
The superconducting ring with two arcs which are characterized by
different critical current densities $j_{c1}<j_{c2}$.}
\end{figure}

This behavior is similar to dependence of the critical current of
superconducting ring formed from two arcs with different critical
currents (see Fig.~\ref{Fig:ring}). When current density in the
upper arc exceeds $j_{c1}$ the phase slip process starts in this
arc and the current density redistributes, leading to $j<j_{c1}$
in the upper arc and nonzero vorticity $N= \oint \nabla \varphi
ds/2\pi$ in the ring. The number of phase slips depends on the
radius of the ring (its length), because each phase slip decreases
the supervelocity and supercurrent on $\sim 1/R\sim 1/L$. In this
system one could expect oscillations of $j_c$  as function of
radius, because with increasing of $R$ vorticity increases
discretely. So the vorticity in such a ring is the analog of phase
soliton in the two-band superconducting bridge.

The amplitude of oscillations of $j_c(L)$ decreases with
increasing constant $\gamma$ and they completely disappear for
sufficiently strong interband coupling
(Fig.~\ref{Fig:jc-coupled}). This occurs due to the suppression of
the independent phase slippage in different bands and absence of
soliton states. But for not very large $\gamma$ finite $\theta(x)$
still exists at the current just below $j_c$ which originates from
the nonuniform distribution of the order parameter along the
bridge. This distribution $\theta(x)$ (so called "phase
texture"~\cite{PhysicaC-2015}) has small amplitude and does not
create finite interband phase difference between the leads.
Similar interband phase distribution also exists for $\gamma\neq
0$ below the crossover to the soliton state (see
Fig.~\ref{Fig:psi-theta}).

\begin{figure}
\includegraphics[width=1.0\linewidth]{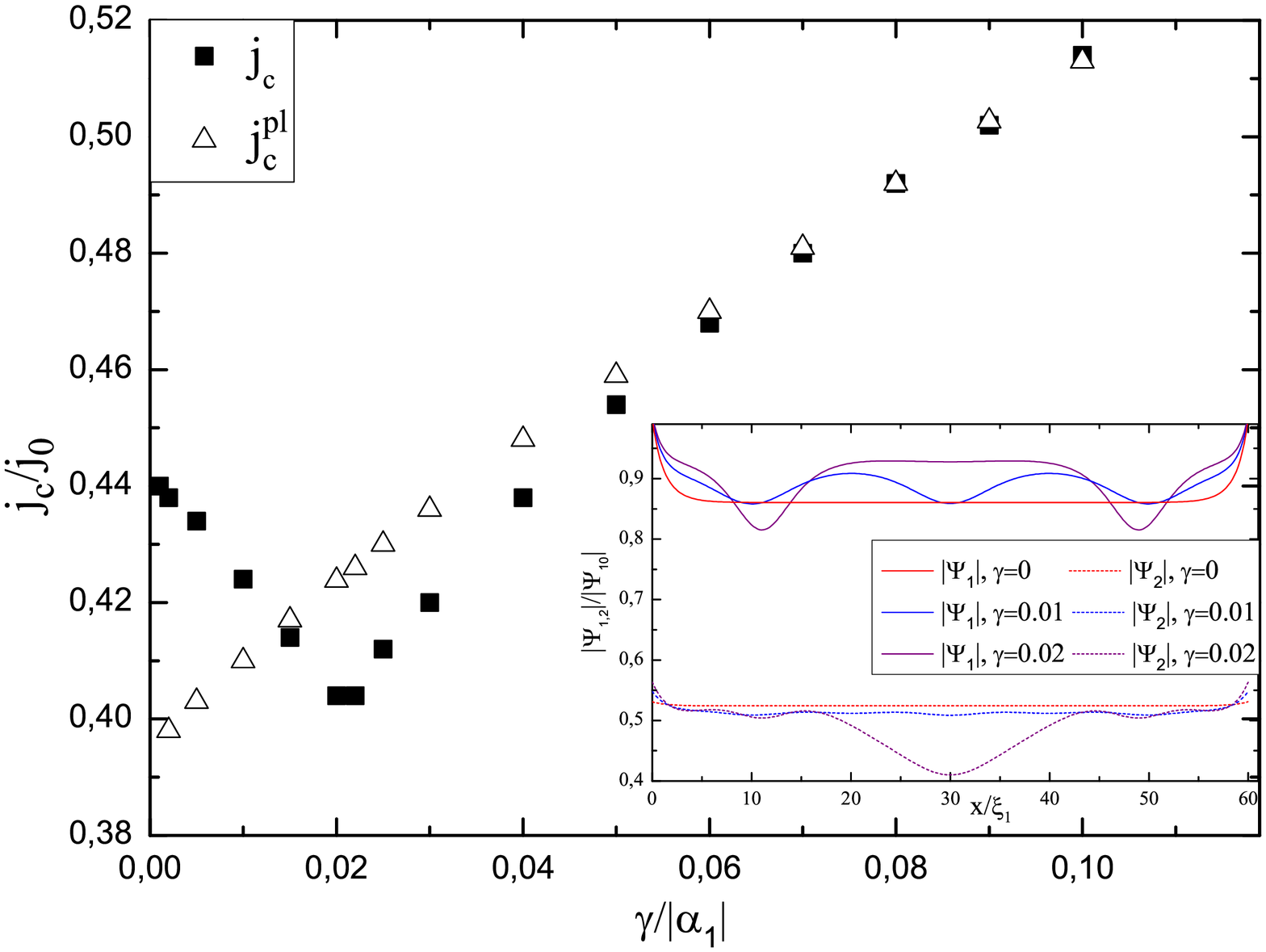}
 \caption{\label{Fig:jc-gamma}
The dependence of critical current density on interband Josephson
coupling $j_c(\gamma)$ for the bridge length $L=60\xi_1$ and
parameters $u=u_1=u_2=0.2$ and $R=0.15$ (squares). Obtained
results are compared with calculated critical current $j^{pl}_c$
in the phase-locked state (triangles). }
\end{figure}

Dependence of $j_c(\gamma)$ is shown for the bridge with
$L=60\xi_1$ in Fig.~\ref{Fig:jc-gamma}. At small $\gamma$ critical
current decreases with increasing $\gamma$ because order parameter
is stronger suppressed in both bands in place of location of phase
solitons \cite{NJP-2012} (see inset in Fig.~\ref{Fig:jc-gamma}),
which makes superconducting stationary state unstable at smaller
current. The minimum of dependence $j_c(\gamma)$ is reached at
$\gamma=0.022$ (for chosen parameters) which corresponds to
transition from phase soliton state to the phase texture state.
With further increase of $\gamma$ critical current increases
because the weaker band is strengthened. In Fig.
~\ref{Fig:jc-gamma} we also plot critical current $j_c^{pl}$ which
is found using the assumption that the transition to the resistive
state occurs from the phase-locked state. \cite{LTP-2007} One can
see that it coincides with our $j_c$ at relatively large $\gamma$
when phase texture state practically disappears (phase texture
state weakens the superconducting state in a way like it does the
solition state) and $j_c<j_c^{pl}$ (excepting rather small
$\gamma$).

We also consider two-band bridge with normal leads. Despite the
presence of phase solitons, the critical current $j_c$ is
practically unaffected by the bridge length, similarly to the case
of zero interband Josephson coupling.

\section{Influence of large relaxation time of $|\Psi_{k}|$ on critical current oscillations}

\begin{figure}
\includegraphics[width=1.0\linewidth]{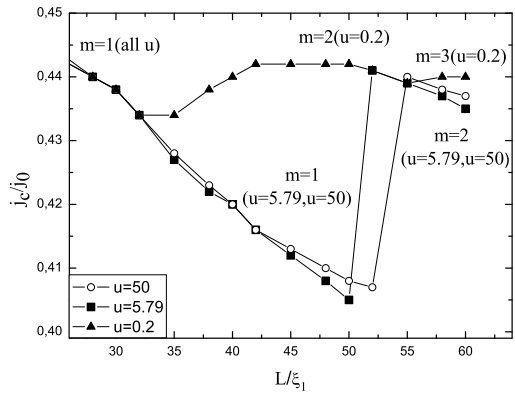}
 \caption{\label{Fig:jc-u}
The dependence of critical current density on a bridge length
$j_c(L)$ in the absence of interband Josephson coupling at
different values of the parameter $u=u_1=u_2$ and $R=0.15$.}
\end{figure}

In our previous calculations we use small value $u_k$. In the case
of superconductors with strong inelastic electron-phonon
scattering $u_k\gg 1$ (Ref.\onlinecite{Vodolazov-2011}) which
affects the phase-slippage process. Therefore we calculate $j_c$
for different lengths, two values of parameter $u=5.79$ and $u=50$
and compare results with case of small $u$ - see
Fig.~\ref{Fig:jc-u}. In the absence of interband Josephson
coupling we find that the critical current depends on parameter
$u$ for sufficiently large lengths, when transition to resistive
state occurs from soliton state. The found effect is connected
with large relaxation time of magnitude of the order parameter
which is proportional to $u$. For large $u$ magnitude of the order
parameter $|\Psi_{1,2}|$ does not have time to recover after phase
slip event and it favors transition to the resistive state instead
of transition to the soliton state with larger $m$. At finite
interband Josephson coupling similar effect of large $u$ also can
be observed, except that the difference starts at larger $L$ and
$m$.

\section{Conclusion}

In the framework of time-dependent Ginzburg-Landau theory we study
critical current of quasi-one-dimensional bridge formed from
two-band superconductor. We found oscillatory dependence of $j_c$
on the length of the bridge which is explained by formation of
phase solitons at $j<j_c$ for relatively long bridges. This effect
is noticeable for relatively small ratio between critical currents
of weak and strong bands (which they would have in absence of
Josephson coupling) and relatively weak Josephson coupling
$\gamma$ between bands.

Most suitable for observation of predicted effect is FeSe$_{0.94}$
($\gamma=0.01$, $T_c=8.3$ K, $T_{c2}=3.1$
K[~\onlinecite{PRL-2010}]) or artificial structure, which is
consisted from two Josephson coupled superconducting layers where
large difference in critical currents could be realized via
creation of artificial defects in the layers. Nonmonotonous
dependence $j_c(L)$ could be seen either via variation of the
length of the bridge or, at fixed length, via variation of the
temperature which leads to change of ratio $L/\xi_1(T)$ (we expect
in this case some kind of kinks on dependence $j_c(T)$).

\begin{acknowledgments}
D.Yu. V. acknowledges support from the Russian Scientific
Foundation for Basic Research, grant No 15-42-02365, and P.M. M.
by grant No. 15-41-02519.
\end{acknowledgments}

\end{document}